\documentclass[prd,twocolumn,superscriptaddress,altaffilletter]{revtex4}
\usepackage{graphicx}
\usepackage{bm}
\begin{document}
\author{Juan M. Aguirregabiria}
\email{wtpagagj@lg.ehu.es}
\affiliation{Fisika Teorikoa, Zientzia eta Teknologiaren Fakultatea, Euskal Herriko Unibertsitatea, 644 Posta Kutxatila, 48080 Bilbao}
\author{Luis P. Chimento}
\email{chimento@df.uba.ar}
\affiliation{
Dpto. de F\'\i sica, Facultad de Ciencias Exactas y Naturales,
Universidad de Buenos Aires, Ciudad Universitaria
Pabell\'on I, 1428 Buenos Aires, Argentina}
\author{Ruth Lazkoz}
\email{wtplasar@lg.ehu.es}
\affiliation{Fisika Teorikoa, Zientzia eta Teknologiaren Fakultatea, Euskal Herriko Unibertsitatea, 644 Posta Kutxatila, 48080 Bilbao}
\title{Phantom k-essence cosmologies}

\begin{abstract}
We devise a method to obtain a phantom version of FRW  k-essence cosmologies with homogeneous k-fields by applying form-invariance transformations. It can be seen that the transformation performs the maps $H\to\-H$ and $\rho+p\to-(\rho+p)$, which in turn give $\gamma\to\-\gamma$ and $a\to a^{-1}$. The discussion is presented in a general setup, valid for FRW k-essence cosmologies, and then we discuss power-law solutions for illustration purposes. First, we deal with  models such that the gradient
of the k-field is not constant, which include standard and generalized tachyon cosmologies. We concentrate on the usual tachyon and show the phantom symmetry involves a change in the potential and that it generates an extended super-accelerated tachyon field. Then, we turn our attention to models for which the time derivative 
of the k-field is not constant, and we show the transformation can be implemented without changing the potential at all.   
\end{abstract}
\maketitle\section{Introduction}
Observational data provided by the WMAP mission \cite{WMAP} seem to have confirmed the existence of an epoch of accelerated expansion in the early universe. In addition, according to observations of distant supernovas \cite{supernovae}, also  the universe at present is expanding with acceleration. The idea that some kinds of scalar fields could be the agents driving those two periods of expansion is widely accepted. Those fields are described by effective theories which, in general, include in their lagrangians non-canonical terms in field derivatives, and might bring in crucial cosmological consequences like the occurrence of inflation even without a potential (purely kinetic acceleration or k-acceleration) \cite{kinflation,scherrer,Chimento}. Given that the equations of motion in all classical theories seem to be of second order, the non-canonical terms considered in the lagrangian will only be combinations of the square of the gradient of the scalar field (hereafter k-field). Moreover, since k-fields can be used for constructing dark energy models it is common place to interpret them as some kind of matter called k-essence   \cite{scherrer,k-essence,tracking}. Note, however, that the  description of late time acceleration was not the original reason why  k-fields were introduced, but rather they were put forward  as possible  inflation driving agents \cite{kinflation,kinfper}. Interestingly enough, as shown in \cite{tracking}, one can also  construct tracking k-essence cosmologies, although there are dynamical systems arguments against their plausibility \cite{against}.

Lately, efforts in the framework of k-essence have been directed towards model building using  power-law solutions \cite{exactsolutions, tacpow}.  Such cosmologies may be interpreted as  universes filled with barotropic fluids with a constant barotropic index.  
In this paper we address  k-essence cosmologies that violate the weak energy condition $\rho>0$, $\rho+p>0$, but from  a  different perspective than earlier works \cite{earlier}. The models will  be dubbed phantom k-essence cosmologies following the terminology in  \cite{phan}.  Phantom matter can apparently be accommodated by current observations \cite{obs}, and even though the theoretical understanding of the acceptability of phantom matter is limited, we can rely on the motivation  provided by string theory \cite{str}. 
Interestingly, the idea that the origin of
dark energy should be searched within a fundamental theory, say string theory, has been recently reinforced by the discovery that the holographic principle cannot be used to tell whether dark energy is present or not \cite{holography}.

Theoretical cosmology with phantom models has become an active area of theoretical research. Sometimes the accent is put on exact solutions
 \cite{construction}, whereas some others cosmological dynamics  is the key subject  \cite{dynamics}. Related to this, at present there is no consensus as to whether a universe that violates the weak energy condition should generically possess a future singularity or big rip \cite{fate}. Now, since the idea of phantom cosmologies is pretty new, even in such a simple setting as (single field) FRW spacetimes many questions  remain open yet. Nevertheless,  not long ago some authors  ventured out of that basic picture, and  pursued generalizations such as considering an AdS geometry \cite{elizalde} or introducing multiple phantom fields \cite{o(n)}.

Our approach to phantom k-essence cosmologies aims at model construction too, following the line of work successfully initiated in \cite{ChiLaz}. As customary, we will assume the k-field is homogeneous. Two different ways to obtain power-law k-essence cosmologies are known, depending on whether the time derivative of the k-field is constant or not. 
In the first case, the scalar field evolves linearly with time and the potential is necessarily of the inverse square form \cite{k-essence}. Power-law tachyon cosmologies  \cite{tacpow} belong to this case. In contrast, 
in the second case, solutions with arbitrary potentials and  non-linear scalar
fields can be found, if one imposes they have a constant barotropic index \cite{exactsolutions}. A nice feature of those solutions is the rich casuistics in the form of the field and its potential for a 
fixed power-law evolution. We apply our symmetry transformations to the two cases, and highlight the differences and similarities between them.

\section{Form-invariance transformations}
We assume our fluid is the source of a spatially flat homogeneous and isotropic
spacetime with line element
\begin{equation}
ds^2=-dt^2+a^2(t)(dx_1^2+dx_2^2+dx_3^2),
\end{equation}
where $a(t)$ is the scale factor, and the expansion or Hubble factor is defined as $H=\dot a/a$. Here and throughout overdoes will denote differentiation with respect to $t$. 

Consider any two different FRW perfect fluid solutions to the Einstein field equations $a$, $\bar a$, each one generated by energy density and pressure $\rho,p$ and $\bar \rho,\bar p$ respectively. The sets of differential equations that have been solved to obtain those solutions are, in fact, different. Now, 
in the framework of a long-term project \cite{form-invariance,ChiLaz}  it has been shown that a link between those cosmological models can be established using a form-invariance transformation which uses as only input the relation between the energy densities of the two fluids. Ours is, therefore, an uncommon equivalence concept. The seed and transformed cosmological models will be characterized by the set of quantities $\{H,\rho,p\}$ and $\{\bar H,\bar \rho,\bar p\}$
which, as usual,  represent the Hubble factor, energy density and pressure. Each set of those  quantities will satisfy the customary Einstein equations.  We will say the second set corresponds to a cosmological model obtained from the seed one through a form-invariance  transformation generated by $\bar\rho(\rho)$. Interestingly, there is one form-invariance transformation \cite{ChiLaz} which preserves the energy density of the fluid and corresponds to
\begin{eqnarray}
&&\bar H=-H,\label{flipone}\\
&&\bar\rho+\bar p=-(\rho+p)\label{fliptwo}.
\end{eqnarray}
Clearly, it flips the sign of the barotropic index $\gamma\equiv-2\dot H/3H^2$, so in what follows we will we referring to it as
to the ``phantom transformation". 
Now, an expanding phantom k-essence cosmology can be constructed if we trade the initial singularity of the $a$ solution  for the final big rip of the $\bar a$ one.

We turn now to the specific setting of k-essence cosmologies with an homogeneous k-field $\phi$ derived
from the lagrangian 
\begin{equation}
{\cal L}=-V(\phi) F(x),
\end{equation}where $x=-\dot\phi^2$. Under these hypotheses,  the k-essence can be interpreted in terms of
a  barotropic perfect fluid  with equation of state $p=(\gamma-1)\rho$. The Einstein equations reduce, then, to 
\begin{eqnarray}
&&3H^2=\frac{VF}{1-\gamma},\label{ein1}\\
&&\dot H=xVF_x\label{ein2}.
\end{eqnarray} It can also be seen that
\begin{equation}
\gamma=-\frac{2xF_x}{F-2xF_x}\label{gamma}.
\end{equation}
A consequence of (\ref{ein1}) and (\ref{ein2}) is the conservation equation
\begin{equation}
(F_x+2xF_{xx})\ddot \phi+3HF_x\dot \phi+\frac{V'}{2V}(F-2xF_x)=0,
\end{equation}
where $V'=dV/d\phi$.

Finally, we get  
\begin{eqnarray}
\bar V\bar F=\frac{1+\gamma}{1-\gamma}\,VF´\label{vflink}
\end{eqnarray}
by applying the phantom transformation  defined by (\ref{flipone}) and (\ref{fliptwo}).

\section{Phantom k-essence cosmologies arising  from symmetries}
In this section we will discuss the application of the phantom transformation to
power-law k-essence cosmologies.

\subsection{Models with $x=\rm{constant}$}
It is well-known that for power-law solutions with $x=\rm{constant}$ the potential is necessary of the form $V=V_0\phi^{-2}$, with $V_0$ a constant. $F$, of course, is also constant. This is the case  in power-law tachyon cosmologies, for instance. In order to avoid unnecessary complications, we will illustrate the phantom symmetry for the usual tachyon case, but, of course, it could be equally applied to other $x=\rm{constant}$ models, like the generalized tachyon cosmologies in \cite{Chimento}.

For the usual tachyon one has
\begin{eqnarray}
&&\rho=\frac{V}{\sqrt{1-\dot{\phi}^2}},\\
&&p=-V{\sqrt{1-\dot{\phi}^2}},\\
&&\gamma=\frac{\rho+p}{\rho}=\dot{\phi}^2\label{gammaphi},
\end{eqnarray}
and the Friedmann equation can be cast as
\begin{equation}
3H^2=\frac{V_0}{\phi^2\sqrt{1-\dot\phi^2}}\label{frie-tach}.
\end{equation}
Applying the phantom symmetry to (\ref{gammaphi}) we get
\begin{eqnarray}\bar\gamma=-\dot\phi^2,\label{gammaphi-phan}\\
3\bar H^2=-\frac{\bar V_0}{\phi^2\sqrt{1+\dot\phi^2}}\label{frie-tach-phan}.
\end{eqnarray}
Thus, actually,  (\ref{frie-tach}) and (\ref{frie-tach-phan}) arise from two different lagrangians. Note also that the sign of the square of the time derivative of the k-field gets reversed in the phantom transformation.

Now,  the requirement that the energy density gets preserved enforces
\begin{eqnarray}
\bar V_0=- V_0\frac{\sqrt{1+{\phi_0}^2}}{\sqrt{1-{\phi_0}^2}}\label{uvecero},
\end{eqnarray}
where we have put $\phi=\phi_0t$ with $\phi_0$ a constant.
Let us see  how this   corresponds to the map $a\to a^{-1}$. Using  (\ref{gammaphi}) and (\ref{frie-tach}) and recalling $a\propto t^{2/3\gamma}$, after some algebra we arrive
at the result
\begin{equation}
\gamma=\frac{2}{\displaystyle 1+\sqrt{1+{9V_0^2}/{4}}}.
\end{equation}
For the transformed solution  we must take $\bar a\propto t^{2/3\bar\gamma}$ and $\bar\phi=\bar\phi_0t$ with $\bar \phi_0$ a constant.  Straightforward calculations which involve (\ref{gammaphi-phan}) and (\ref{frie-tach-phan}) give 
\begin{equation}
\bar\gamma=\frac{2}{\displaystyle 1-\sqrt{1+{9\bar V_0^2}/{4}}} \label{gammabar}.
\end{equation}
Finally,  if we insert (\ref{uvecero})    back into  (\ref{gammabar})
and do some more algebra, we  see $\bar\gamma=-\gamma$.

\subsection{Models with $x\ne\rm{constant}$}
Let us first review how power-law models are obtained in this case. The conservation equation for $\gamma=\rm{constant}$ can be readily integrated to give
\begin{equation}
VF=\frac{\rho_0}{a^{3\gamma}},\label{firstint}
\end{equation}
where $a\propto t^{2/3\gamma}$ and $\rho_0$ a constant. If (\ref{gamma}) is viewed as a differential equation for $F(x)$, one can solve to yield 
\begin{equation}
F(x)=(-x)^{\gamma/2(\gamma-1)}\label{formof}.
\end{equation}
 Note that in this case we would not have to make any requirement 
on the form of $V$, unless we wished to obtain  $\phi(t)$ explicitly on using (\ref{firstint}) and (\ref{formof}).
Nevertheless,  any $F$ like (\ref{formof}) will give a power-law solution for every potential. 

Let us assume now that $V$ does not change ($\bar V=V$), because  power-law solutions in this case are obtained without making any assumption on $V$, as we just showed. We are going to deduce now the $\bar x (x)$ rule necessary to implement the phantom symmetry. Since we are looking for new power-law solutions, it is admissible, with the requirement of form-invariance, to assume from the beginning
\begin{equation}
\bar F= (-\bar x)^{{\bar\gamma}/2({\bar\gamma-1})}. \label{ftrans}
\end{equation}
Combining (\ref{ftrans}) with (\ref{vflink}), and using  $\bar\gamma=-\gamma$,  we get 
\begin{equation}
{\dot{\bar\phi}}\,^2=\left(\frac{1+\gamma}{1-\gamma}\right)^{2(\gamma+1)/\gamma} \,{\dot{\phi}}^{2(\gamma+1)/(\gamma-1)}\label{xbarra}
\end{equation}
The latter is, in fact, a very interesting result. In standard scalar field theories,
the phantom transformation can be implemented just by Wick rotating the field \cite{ChiLaz}, and
that was also the case in tachyon cosmology, as shown in the previous section. 
In contrast, in the $x\ne\rm{constant}$ case, the transformation rule for the field, as given 
by (\ref{xbarra}), is not so simple and does not include the Wick rotation as a particular case either.

\section{Conclusions}
As part of a long-term project \cite{ChiLaz,form-invariance} we have shown here that form-invariance transformations can be used as tools for generating new exact solutions to the Einstein field equations. In particular, we have applied the method to the obtention of phantom versions of FRW  k-essence cosmologies, with an accent on power-law spacetimes. The discussion has been presented in a general setup, valid for  FRW  k-essence cosmologies, and we have only discussed power-law models for illustration purposes.

 Specifically, we have been concerned with  two families of such solutions, namely, those corresponding to a scalar field (k-field) with constant and non-constant time derivative  respectively. In broad terms, it can be seen that the transformation flips the sign of both the barotropic index and the Hubble factor. Then, if  the initial singularity of the seed solution is identified with  the final big rip of the transformed one,  an expanding phantom universe is obtained.  
 
 Interestingly enough, we have shown that in the $x=\rm{constant}$ cases implementation of the phantom transformation  requires a change in the potential. In the particular case of the tachyon, a sign reversal in the square of the time derivative of the scalar field is also needed; not surprisingly, perhaps, this is exactly the same rule as for standard  scalar field cosmologies. 
Note also that the usual tachyon is a k-essence model with $F=\sqrt{1+x}$, whereas the phantom tachyon  falls into a different category within k-essence models, because it corresponds to $F=\sqrt{1-x}$.  In contrast,   in the $x\ne\rm{constant}$ case, the phantom transformation  can be realized without changing the potential, but the transformation rule for the scalar field is not as simple as  the previous case.
 
Summarizing, we have given a neat prescription for generating phantom k-essence cosmologies,
and we have shown that, against what one can naively expect, in some cases a simple ${\dot \phi}^2\to -{\dot \phi}^2$ map does not do the job. Nevertheless, we believe the subject deserves further investigation, and even though we have concentrated here on power-law spacetimes, hopefully we will   widen our scope in the future to address other cases. 
\section*{ACKNOWLEDGMENTS}
We are grateful to Alexander Feinstein and Alberto D\'\i ez Tejedor for conversations.
LPC is partially funded by the University of Buenos Aires  under
project X223, and the Consejo Nacional de Investigaciones Cient\'{\i}ficas y
T\'ecnicas. JMA and RL are financially endorsed by the University of the Basque Country through research grant 
UPV00172.310-14456/2002. JMA also acknowledges support from the Spanish Ministry of Science and Technology through research grant  BFM2000-0018. RL is also is supported by the Basque Government through fellowship BFI01.412, the Spanish Ministry of Science and Technology
jointly with FEDER funds through research grant  BFM2001-0988. 

\end{document}